# Charge density wave with suppressed long-range structural modulation in canted antiferromagnetic kagome FeGe

Chenfei Shi[1#], Wenchang Hou[2,3#], Hanbin Deng[4], Bikash Patra[5], Surya Rohith Kotla[6], Yi Liu[7, 8], Sitaram Ramakrishnan[9], Claudio Eisele[6], Harshit Agarwal[10,11], Leila Noohinejad[12], Ji-Yong Liu[13], Tianyu Yang[4], Guowei Liu[4], Bishal Baran Maity[5], Qi Wang[3, 14], Zhaodi Lin[1], Baojuan Kang[1], Wanting Yang[1], Yongchang Li[1], Zhihua Yang[15], Yuxiang Chen[16], Xiang Li[16], Yuke Li[15], Yanpeng Qi[3, 14, 17], Arumugam Thamizhavel[5], Wei Ren[1, 18], Guang-Han Cao[8, 19], Jia-Xin Yin[4], Bahadur Singh[5*], Xuerong Liu[2, 3*], Sander van Smaalen[6*], Shixun Cao[1, 18*], and Jin-Ke Bao[1, 15, 18*]

*[1]Materials Genome Institute, Department of Physics, Institute for Quantum Science and Technology, Shanghai University, Shanghai 200444, China*

*[2]Center for Transformative Science, ShanghaiTech University, Shanghai 201210, China*

*[3]School of Physical Science and Technology, ShanghaiTech University, Shanghai 201210, China*

*[4]Department of Physics, Southern University of Science and Technology, Shenzhen 518055, Guangdong, China*

*[5]Department of Condensed Matter Physics and Materials Science, Tata Institute of Fundamental Research, Mumbai 400005, India*

*[6]Laboratory of Crystallography, University of Bayreuth, Bayreuth 95447, Germany*

*[7]Key Laboratory of Quantum Precision Measurement of Zhejiang Province, School of Physics, Zhejiang University of Technology, Hangzhou 310023, China*

*[8]School of Physics, Zhejiang Province Key Laboratory of Quantum Technology and Devices, Zhejiang University, Hangzhou 310027, China*

*[9]Institut NÉEL, CNRS, Univ. Grenoble Alpes, Grenoble 38000, France*

*[10]Laboratory of Crystallography, Bayerisches Geoinstitut, University of Bayreuth, Bayreuth 95440, Germany*

*[11]Institut für Physik, Johannes-Gutenberg-Universität Mainz, Mainz 55128, Germany*

*[12]P24, PETRA III, Deutsches Elektronen-Synchrotron DESY, Hamburg 22607, Germany*

*[13]Department of Chemistry, Zhejiang University, Hangzhou 310058, China*

*[14]ShanghaiTech Laboratory for Topological Physics, ShanghaiTech University, Shanghai 201210, China*

*[15]School of Physics and Hangzhou Key Laboratory of Quantum Matters, Hangzhou Normal University, Hangzhou 311121, China*

*[16]Centre for Quantum Physics, Key Laboratory of Advanced Optoelectronic Quantum Architecture and Measurement (MOE), School of Physics, Beijing Institute of Technology, Beijing, 100081, China.*

*[17]Shanghai Key Laboratory of High-resolution Electron Microscopy and ShanghaiTech Laboratory for Topological Physics, ShanghaiTech University, Shanghai 201210, China*

*[18]Shanghai Key Laboratory of High Temperature Superconductors, Shanghai University, Shanghai 200444, China*

[19]Collaborative Innovation Centre of Advanced Microstructures, Nanjing University, Nanjing 210093, China

# Chenfei Shi and Wenchang Hou contribute equally to this work.
E-mail address: bahadur.singh@tifr.res.in (Bahadur Singh)
liuxr@shanghaitech.edu.cn (Xuerong Liu); smash@uni-bayreuth.de (Sander van Smaalen);
sxcao@shu.edu.cn (Shixun Cao); baojk7139@gmail.com (Jin-Ke Bao)

**Abstract**: Kagome lattice can host abundant exotic quantum states such as superconductivity and charge density wave (CDW). Recently, successive orders of A-type antiferromagnetism (AFM), CDW and canted AFM have been manifested upon cooling in kagome FeGe. However, the mechanism of CDW and interaction with magnetism remains unclear. Here we investigate the evolution of CDW with temperature across the canted AFM by single-crystal x-ray diffraction, scanning tunneling microscope (STM) and resonant elastic x-ray scattering (REXS). Interestingly, CDW-induced superlattice reflections become weak after the canted AFM, although long-range CDW order is still detectable by STM and REXS. We uncover a novel long-range CDW order with suppressed structural modulation, likely due to the competition for the underlying crystal structure between CDW and canted AFM. Additionally, occupational modulations of Ge1 in the kagome plane and displacive modulations of all atoms were extracted. The results confirm Ge dimerization along the *c* axis and suggest a dynamic transformation between different CDW domains.

## Introduction

Over the past decades, strongly correlated electron systems, in which Coulomb repulsive interactions between electrons cannot simply be described as a perturbation, manifest a grand challenge in unfolding the mechanism that determines their intriguing and intricate properties beyond the picture of a non-interaction system[1-4]. Complexity is a ubiquitous characteristic in these systems due to the interactions between different degrees of freedom involving charge, spin, lattice and orbital, as exemplified by high-$T_c$ superconductors[5-7], colossal-magnetoresistance manganites[8,9], heavy Fermion compounds[10], two-dimensional Moiré systems[11,12] and organic conductors[13,14]. A plethora of novel quantum states such as superconductivity, charge/spin density waves[15,16], exciton condensation[17] and Wigner crystallization[18] emerge from the extensive parameter space covered by these degrees of freedom. As a result, strongly correlated electron systems provide a fertile playground to study the competition and/or intertwinement of such quantum states and further manipulate them by controlling external parameters such as carrier concentration, temperature and pressure. These researches will assist in establishing an emerging paradigm for strongly correlated electron systems.

The kagome lattice, a two-dimensional network of corner-shared triangles with geometric frustrations, exhibits the characteristics of flat bands, van Hove singularities (vHSs) and Dirac dispersion in its electronic structure[19]. It can produce strong electron correlations due to the quenching of kinetical energy by quantum interference from its special geometry, electronic instability from high density of states as well as topological properties induced by spin-orbit coupling from a massless band[20]. Thus, materials with a kagome lattice serve as an excellent platform to research on interesting quantum phenomena. For instance, the mineral Herbertsmithite with a kagome lattice of $Cu^{2+}$ ions is proposed to be the long-sought quantum spin liquid due to the highly frustrated antiferromagnetic (AFM) interactions[21]; $AV_3Sb_5$ ($A$ = K, Rb and Cs) with a kagome lattice of V atoms exhibit various quantum states such as superconductivity with a pair density wave order ($T_C \approx 0.92$ - 2.5 K), time-reversal-symmetry-breaking charge density wave (CDW) ($T_{CDW} \approx 78$ - 103 K) and $Z_2$ topological states[22] while their isostructural compound $CsCr_3Sb_5$ has been unveiled to undergo concurrent CDW and spin density wave orders ($T \approx 55$ K) which can be suppressed to realize superconductivity under high pressure[23]. Thus, the investigations of kagome materials with significant electron correlations can further shape the research paradigm in strongly correlated electron systems.

Recently, B35-type FeGe with the kagome lattice of Fe atoms presents a cascade of quantum orders with successive transitions toward an A-type AFM order with magnetic moments perpendicular to the kagome plane at $T_N \approx 410$ K, a 2 × 2 × 2 supercell short-range CDW order at $T_{CDW} \approx 100$ K and a double-cone AFM spin canting transition at $T_{canting} \approx 60$ K[24,25]. The short-range CDW order in FeGe can be tuned into a long-range order by post-annealing treatments[26-28]. The rich quantum phenomena such as CDW and topological edge states have stimulated intensive researches on this system both theoretically and experimentally[29-34]. The major CDW-induced structural distortion comes from the dimerization of Ge sites located in the kagome plane along the *c* axis as supported by x-ray diffraction[26-28,35], angle-resolved photoemission spectroscopy[36,37] and DFT calculations[28,34,35]. The magnetic moment of Fe shows a slight increase when it enters the CDW state, indicating a strong coupling between AFM and CDW orders. Its sizable magnetic moment ($m_{Fe} \approx 1.7\ \mu_B$) for AFM[38,39] in FeGe puts it in a category of strongly correlated electron systems while $AV_3Sb_5$ are non-magnetic with weak electron correlations[40]. CDW might be susceptible to and even suppressed by a ferromagnetic order as observed in colossal-magnetoresistance manganites[41] and Sm(Nd)$NiC_2$[42,43]. As for FeGe, understanding the interactions between CDW and magnetism will provide more hints about the mechanism of the novel CDW which is still under an intensive debate[31,34,35,44]. CDW fluctuations above $T_{CDW}$ and below $T_N$ have also been identified by diffuse x-ray scattering signal and the existence of a tiny fraction of Ge dimerization in FeGe[26]. However, the explicit evolution of its CDW order below $T_{CDW}$, especially across spin canting transition, still needs to be unveiled to outline the complete physical picture for this system.

Here, we investigate detailed structural distortions from the CDW order with temperature by single-crystal x-ray diffraction (SXRD) and carry out a (3+3)-dimensional commensurate structure refinement analysis in the framework of superspace on the modulated structure in FeGe. Dynamic exchange of different CDW domains from two distortion modes along the *c* axis with temperature is obtained from the structure refinements. Strong superlattice reflections from structural modulation become much weaker abruptly below the spin-canting transition while the robustness of CDW order in FeGe is corroborated by scanning tunneling microscope (STM) and resonant elastic x-ray scattering (REXS) measurements.

## Results and Discussions

### Superlattice-reflection melting

In order to study more explicitly the evolution of the CDW order with temperature in FeGe, single crystal x-ray diffraction was performed at several temperatures covering the ranges above and below $T_{CDW}$ as well as $T_{canting}$. Several reconstructed layers in the reciprocal space are shown in Fig. 1 for 270, 110 and 20 K. Firstly, weak signals from some superlattice reflections in the *ab* plane have been already identified at the (*hk*2) diffraction plane above $T_{CDW}$ at 270 K (Fig. 1(a) and Fig. 1(g)), indicating the existence of short-range CDW correlations within *ab* plane[26]. However, no diffracting signal from the doubling of the *c* axis for the $2 \times 2 \times 2$ CDW can be identified at 270 K, see the reciprocal images of (0*kl*) plane at 270 K (Fig. 1(d)). These observations suggest that the CDW coherence length along the *c* axis is much smaller than that in the *ab* plane above $T_{CDW}$. Strong superlattice reflections appear at 110 K below $T_{CDW}$, forming a long-range ordered $2 \times 2 \times 2$ CDW, see the (*hk*2) and (0*kl*) planes in Fig. 1(b), (e) and (h). They can be indexed by three independent **q**-vectors: (0.5, 0, 0), (–0.5, 0.5, 0) and (0, 0, 0.5), which are consistent with previous reports[26,35]. The behaviors at 80 and 60 K are more or less the same to the case at 110 K, showing strong signals of superlattice reflections (Fig. S1 in Supporting Information). Surprisingly, the number of observable superlattice peaks decreases significantly at 20 K, see Fig. 1(c), (f) and (i). The superlattice reflections related to the in-plane **q**-vectors of (0.5, 0, 0) and (–0.5, 0.5, 0) can only be observed at low angles, see the (*hk*2) plane in Fig. 1(c), and almost no reflections related to **q**-vector of (0, 0, 0.5) are observed, see the (0*kl*) plane in Fig. 1(f). The intensities of superlattice reflections at 20 K are indeed much weaker than that at 110 K by doing line cuts in the planes (Fig. S2 in Supporting Information). The FWHM at 20 K is also larger than the one at 110 K. The case at 40 K is similar to that at 20 K with significantly less pronounced signals of superlattice reflections (Fig. S1 in Supporting Information).

These results are in stark contrast to the reports from other groups[24,27,35,45]. They don't observe a significant decrease of the intensity for the superlattice reflections from either neutron or x-ray diffractions below $T_{canting}$. We notice that their crystals used for these measurements are either as-grown or annealed at 593 K for 4 days while the data in our work are collected from the crystals annealed at 573 K for 10 days. We also observed similar results from the crystals annealed at 643 K for 10 days. In order to confirm that annealing periods are also crucial in determining the behavior below $T_{canting}$, we performed low temperature x-ray diffraction on the crystals annealed at 573 K for 2 and 4 days, shorter periods than that of the crystal used in Fig.1. Despite a sharp drop in the magnetic susceptibility and strong superlattice reflections just below $T_{CDW}$, no significant suppression of the superlattice reflections below $T_{canting}$ is identified, see Fig. S3 in Supporting Information, consistent with the reports from other groups[24,27,35,45].

The weak superlattice reflections originating from short-range CDW in the as-grown sample #B also vanished when the sample was cooled down to 20 K (Fig. S4 in Supporting Information). As for another as-grown sample #C which does not exhibit long-range CDW order, the intensities of superlattice peaks do not have a significant reduction at 20 K compared with the case at 75 K in the CDW state (Fig. S5-6 in Supporting Information). The melting of superlattice reflections from short-range CDW order at low temperatures can be somehow avoided in some as-grown samples and the annealed sample with a shorter period because of the defect pinning of the structural modulation[45], which is why it is not observed in the original report and other works in FeGe[24,35], while it always occurs in our samples annealed for a longer period with a long-range CDW order, suggesting an intrinsic property for a high-quality sample with less disorder. As a result, the defect levels plays an essential role in determining not only the bulk nature of CDW order but also the structural modulation below $T_{canting}$. More future study are needed to elucidate the actual mechanism for this novel behavior.

In order to resolve the temperature at which the melting of superlattice reflections begins, we tracked temperature-dependent integrated intensities of two superlattice reflections (1.5 –2 2) and (1 –1.5 2) which are normalized by the main reflection (1 –1 2), see Fig. 2(b). Their intensities dropped suddenly to a small value just below 60 K, close to the spin canting transition determined by magnetic susceptibility (Fig. 2(a)), and remained to be a small value instead of dropping to zero, consistent with the reciprocal images at 20 K (Fig. 1(c), (f) and (i)). When the sample was warmed up across the spin canting transition to 75 K, the signals of the superlattice peaks returned back to the condition at 110 K (Fig. S7 in Supporting Information). The suppression is not due to the radiation damage of the sample during the measurements but an intrinsic phenomenon. The lattice parameters of *a* and *c* also exhibit a subtle increase below the spin canting transition (Fig. 2(d)), pointing to a magnetostriction effect in FeGe[46]. The reentrance of superlattice-reflection melting below $T_{canting}$ in FeGe is also supported by temperature-dependent Raman modes with a *RR* scattering geometry by Wu *et al.*[47] and the band shift in the energy distribution curves in ARPES results by Oh *et al.*[36], both of which exhibit the reentrant behavior, suggesting a competing scenario between CDW and other electronic orders[41,42,48,49]. Long-range CDW order below spin canting transition survives as shown below by STM and REXS in spite of the loss of its induced long-range structural modulation. Indeed, previous DFT calculations on the $2 \times 2 \times 2$ supercell with a canted AFM do not favor a centrosymmetric structure with space group *P*6/*mmm* in terms of energy[34] while CDW-induced distorted structure is revealed to be *P*6/*mmm* by both experimentally and theoretically[26,34]. We also carried out DFT calculations on the A-type AFM and simple canted AFM configurations to check their total energies. We found out that appropriate dimerization of Ge atoms in the $2 \times 2 \times 2$ supercell has a lower energy than the non-dimerization case under the A-type AFM, consistent with previous calculations[50], see Fig. S8(a) in Supporting Information. However, a simple canted AFM does not change the overall energy landscape for the dimerization process, see Fig. S8(b) in Supporting Information, suggesting that the current calculations does not confirm the non-dimerization structure below $T_{\mathrm{canting}}$. One probable reason is due to the inaccurate description of the low-temperature canted AFM model which does not fully match the recent neutron diffraction data[45]. Instead, a possible spin density wave component should be considered for the low temperature phase[51]. Another possible reason is the existing disorder in the sample, which can modify the energy landscapes for those two structures. However, such a random disorder is difficult to model for an actual sample[52]. More future study is needed to elucidate a scenario of the competition of CDW and canted AFM in determining the ultimate underlying crystal structure below $T_{\mathrm{canting}}$.

**Modulated structure in (3+3)-dimensional superspace**

Although a commensurate structure with a $2 \times 2 \times 2$ supercell can be refined by a regular space group, the superspace group method provides a concise way to concentrate on the structural distortion from CDW because the refinement is performed by adding additional **q**-vector-dependent harmonic functions to the average structure of the original unit cell such that the most significant distortion of the modulated structure can be easily grasped[53,54]. There are three independent **q**-vectors as mentioned above to index all the superlattice reflections in FeGe, suggesting a (3+3)-dimensional modulated structure. The results with occupancy modulations as well as the displacements of all the atoms for 110 K are presented as an example in the main text (Fig. 3-4) and the information of interatomic distances are also given in Supporting Information (Fig. S9-10).

Since commensurate modulated structures only contain segmental points of the whole modulated functions, we made line cuts with the continuable variable *t* and kept the other two variables *u* and *v* as possible fixed values to present the modulation amplitudes, see Fig. 3 and 4. There are large occupational modulations for Ge1_1 (between 0.063 and 0.98) and Ge1_2 (between 0.93 and 0.0051) under $u = 0$ and $v = 0.25$ due to the partial dimerization of Ge1 atoms, see Fig. 3(a). For an ideal case with one single CDW domain, the occupancy should be either 0 or 1 in the commensurate cuts of the modulation functions. This causes 1/4 of Ge1 atoms located in the kagome plane to form a dimer with Ge-Ge distances modulated from 2.67 to 5.41 Å along the *c* axis (Fig. 6(b)). However, different CDW domains with a possible $\pi$ phase shift along three doubled axes can coexist due to possible crystal defects acting as the domain walls, which leads to structural disorder in the refinement on the diffracted intensity from all the CDW domains. As for $u$ = 0 or 0.5 and $v = 0.75$, the occupation modulations of Ge1_2 atoms are small, see Fig. 3(b), indicating that the volume of other domains for the dimerization along the *c* axis with a $\pi$ phase shift is quite small at 110 K.

The average occupancy fraction for the Ge1_1 atom without undergoing the dimerization process below $T_{\mathrm{CDW}}$ and above $T_{\mathrm{canting}}$ keeps almost the ideal value of 0.75 (Fig. 2(c)) for a $2 \times 2 \times 2$ CDW supercell where 1/4 of the Ge atoms in the kagome plane exhibit the dimerization[26,28,34], corroborating an almost bulk nature of the CDW ordering. The average occupancy of the Ge1_2 atom related to the distortion mode 1 (see the red arrows in Fig. 7(b)) decreases with temperatures below $T_{\mathrm{CDW}}$ while that related to the distortion mode 2 (see the purple arrows in Fig. 7(b)) increases, suggesting a dynamic volume transformation between domains with temperature (Fig. 2(c)). Above $T_{\mathrm{CDW},}$ the volumes contributing these two modes are equal but small. The displacive modulations for Ge1_1 and Ge1_2 atoms are only along the *z* (the same to *c*) axis (Fig. 3(c) and (d)). There are no occupational or displacive modulations for Ge1_1 and Ge1_2 atoms with $u$ = 0.5 because the generated atoms along the *t* direction are symmetry-related by a six-fold rotation along the *c* axis. The displacive modulations of Ge1_1 for *v* = 0.25 and 0.75 are also symmetry-related by a mirror perpendicular to the *c* axis. As for Ge2 atoms, the displacive modulations are in the *ab* plane with opposite directions for *v* = 0.25 and 0.75, indicating a $\pi$ phase shift between neighboring Ge honeycomb planes (Fig. 4(a) and (b)). The Ge2 atoms generated by the high superspace cut (*t*, *u*, *v*) = (0.5, 0.5, 0.25) or (0.5, 0.5, 0.75) are located in a high-symmetry position with a fixed coordinate, corresponding to the zero displacement in Fig. 4(b). The dominant displacements of Fe are along the *c* axis while its in-plane displacements in all the commensurate cuts are tiny but non-zero (d*x* = 0.00072 Å, d*y* = 0.00149 Å for the cut with the smaller value) (Fig. 4(c) and (d)) which has not been captured experimentally[26] before due to the limited accuracy. There is a mirror symmetry related to the Fe kagome planes located at *v* = 0.25 and 0.75.

The Fe-Fe distances in the kagome planes are modulated from 2.48 to 2.50 Å (Fig. S9 in Supporting Information) while Ge-Ge distances in the honeycomb planes are modulated from 2.80 to 2.96 Å (Fig. S10 in Supporting Information) at 110 K, which are much smaller than the Ge-Ge dimerization modulation along the *c* axis (Fig. S11 in Supporting Information). The modulated structure at 80 K is similar to 110 K with reduced displacement amplitudes (Fig. S12-13 in Supporting Information).

The superlattice reflections at 20 and 40 K are too weak to perform a reliable (3+3)-dimensional modulation refinement. Instead, the periodic average structure was refined, while disregarding any superlattice reflection (Table S13-S14 in Supporting Information). The average occupancy of Ge1_1 is around 0.954 and 0.940 at 20 and 40 K, respectively, significantly higher than 0.75 at 60-110 K and close to 0.951 at 270 K, proving a reentrant short-range structural modulation (Fig. 2(c)). The average occupancy of Ge1_1 is around 0.837 at 20 K for the as-grown sample #C and shows almost no difference compared to the value at 75 K, indicating a possible pinning of modulated structure with little change in the intensities of superlattice reflections (Fig. S4-5 in Supporting Information).

**CDW with suppressed structural modulation**

Since no obvious anomaly has been identified in resistivity and the electronic structure does not exhibit significant reconstructions but a band shift across the spin canting transition in FeGe[26,28,36,37], CDW probably survives although the long-range modulated structure is suppressed. To further confirm such a scenario, STM measurements probing the electronic states directly near to Fermi level below $T_{canting}$ was carried out. The topographic image mainly reflecting the distribution of atoms, i.e. the lattice structure, shows no obvious sign of $2 \times 2$ supercell in a large area at 4.7 K (Fig. 5(a)), as also evidenced by its Fourier transform with extremely weak $2 \times 2$ superlattice peaks (Fig. 5(b)). This is consistent with the melting of superlattice reflections in the x-ray diffraction below $T_{canting}$. However, the difference conductance d*I*/d*V* map mainly reflecting the density of states at Fermi level on the exactly same region as the tomographic image exhibits a pattern with a larger periodicity (Fig. 5(c)), which is clearly demonstrated by the Fourier transform with $2 \times 2$ charge modulation of in-plane triple-**q** vectors (Fig. 5(d)). Even higher-order **q**-vectors with interferences among them have been observed (Fig. 5(d)), further proving that the $2 \times 2$ charge modulation at 4.7 K below $T_{canting}$ is quite strong and robust. The intensity distribution of the peaks from three different **q**-vectors forms a possible chirality, breaking the six-fold rotation symmetry although the refined crystal structure still remains such a symmetry. The weak superlattice peaks in Fig. 5(b) might come from (1) the contribution of charge modulation since the data of tomography will inevitably convolute both the atomic structure and its density of states or (2) trapped regimes with $2 \times 2$ structural modulation. Our results are not contradictory to an STM study by Chen *et al.* in which strong $2 \times 2$ charge modulation in the d*I*/d*V* map is identified while much weaker $2 \times 2$ superlattice intensity signal from Fourier transform of the tomographic image is observed[28]. The weak superlattice in their study is due to either the possible pinning effects of structural modulation as also the case in our annealed samples below $T_{canting}$ or the convolution from the strong modulation of electronic density of states in the d*I*/d*V*. Besides, the bias voltage (60 meV) in our tomographic measurement here is much smaller than them (200 meV), which will include less contribution of the electronic density of states.

Since STM is a surface-sensitive technique, the detected charge modulation could be a surface phenomenon only. To further confirm the bulk nature of the CDW below $T_{canting}$ in FeGe, we performed the REXS measurements with a photon energy of around 11 keV which will penetrate through the sample. Fig.6 shows temperature-dependent intensity of the peak with a moment transfer **q** = (2.5 0 1). Its intensity is close to zero at 120 K above $T_{\mathrm{CDW}}$ and begins to rise at around 110 K corresponding to the CDW transition. As expected, it starts to drop at around 60 K due to the suppression of the structural modulation as shown above. The intensity remains finite at 20 K as demonstrated by the K-scan of the peak at different temperatures in Fig. 6(b). The two neighboring shoulders near the main peak are due to small portion of slightly misaligned intergrown crystals. Such a signal comes from the contribution of charge modulation of valence electrons even though long-range structural modulation is suppressed. To further verify that the CDW is indeed long-range ordered below $T_{\mathrm{canting}}$, temperature-dependent FWHM of the same peak **q** = (2.5 0 1) in a K-scan mode can be extracted by a Gaussian fit, see Fig. 6(c). The value of FWHM is 0.0023 r.l.u. (reciprocal lattice unit) at 100 K when entering the long-ranged CDW state and shows a steady decrease below 100 K with almost no change during the spin canting transition. The value of FWHM is 0.00215 r.l.u. at $T_{\mathrm{canting}}$ and is 0.0021 r.l.u. at 25 K. However, the estimated FWHM has a significant increase for the residual superlattice peaks observed by x-ray diffraction (Fig. S2). This suggests that the long-range CDW persists while the long-range structural modulation is suppressed, consistent with the recent STM[28] and ARPES results[36,37].

**Temperature-dependent phase diagram**

Based on the above results, the cascades of quantum states and their crystal structures with temperature for annealed FeGe crystals are summarized in Fig. 7 where the arrow from left to right indicates a gradual decrease of temperatures. The short-range CDW order inferred from short-range structural modulation starts to be present below $T_{\mathrm{N}} \approx$ 410 K (see the refinement results at 400 K in Table S15-16 in Supporting Information), and grows into a long-range CDW order accompanied by a long-range structural modulation of $2 \times 2 \times 2$ supercell below $T_{\mathrm{CDW}} \approx 110$ K. Interestingly, the long-range structural modulation is suppressed into a short-ranged one but the long-range CDW order still survives when the A-type AFM along the *c* axis transforms into a double-cone canted AFM below $T_{\mathrm{canting}}$, suggesting the competition between these two types of order in determining the actual structure. 270, 110 and 20 K are used as three typical temperatures to represent three different regimes in FeGe. The crystal structure for long-range structural modulation at 110 K exhibits a dominant Ge1_1 dimerization with a displacement of 0.72 Å along the *c* axis and subtle structural distortions on the Fe kagome and Ge honeycomb (Kekulé pattern) planes. one fourth of the total Ge1_1 sites in the kagome plane form the dimerization in the $2 \times 2 \times 2$ supercell at 110 K. At 270 and 20 K, a tiny fraction of Ge1_1 dimerization does not lead to a long-range $2 \times 2 \times 2$ supercell but shows short-range structural modulation as reflected by disorder in the average crystal structure. Only 0.046 and 0.048 of the total Ge1_1 sites exhibit the dimerization process at 270 and 20 K, respectively, which are far from 0.25 in a long-range structural modulation. The space group *P*6/*mmm* is the best option to describe the crystal structure regardless of short-range or long-range structural modulations. There is also a possibility that CDW-induced long-range structural modulation still exists below $T_{canting}$ but its amplitude is so weak to be detected by the current synchrotron radiation source. Our data do not underpin any lower-symmetry models such as monoclinic or orthorhombic symmetry as observed in a recent study on FeGe[47], probably because those lattice distortions are too weak to be detected with the current resolution of the equipment or the samples used in our work are distinct from theirs in terms of defect levels.

## Conclusion

The CDW-induced long-range structural modulation below $T_{CDW}$ is suppressed in annealed crystals of FeGe when it enters a canted AFM state in which only short-range structural modulation exists but long-range CDW order survives. The suppression of superlattice reflections from structural modulation in the as-grown samples depends on the condition of each crystal, suggesting that defects in crystals might pin the CDW-induced structural modulation as observed in a recent low-temperature scanning-TEM study[45]. Structure refinements in a (3+3)-dimensional superspace unveiled that the dominant distortions for the CDW-induced modulated structure are displacements along the *c* axis of Ge1 atoms located in the Fe kagome planes, while the other atoms have much smaller displacements. The extremely tiny in-plane displacement for all the Fe atoms in the kagome plane is also obtained from such refinements. Domains with two different dimerization modes in the CDW state above $T_{canting}$ change their volume ratios with temperature, indicating a dynamic transformation between them. Our work demonstrated a possible state with coexistence of CDW and canted AFM orders but without long-range structural modulation in stark contrast to a common CDW order with a significant periodic lattice distortion accompanied[55], which can serve as a canonical example to study the intricate interactions between charge, spin and lattice in strongly correlated electron systems.

## Methods

### Single crystal growth and physical property measurements

Single crystals of B35-type FeGe were synthesized by using the same procedures as in Ref.[26]. The crystals presenting long-range CDW order were annealed at 573 or 643 K for 10 days. Crystals are also annealed at 573 K for 2 or 4 days for a comparison. The temperature-dependent magnetic susceptibility was measured under $H \perp c$ ($\mu_0 H$ = 1 T), using a commercial superconducting quantum interferometer (MPMS3, Quantum Design). The crystal measured by x-ray diffraction exhibits a CDW transition at around 113 K and a spin canting transition at around 60 K (Fig. 2(a)). The detailed physical property measurement for as-grown sample #B and #C as well as the annealing sample have been reported by us before and can be traced back in Ref.[26]. STM measurements about the tomographic image and its corresponding d$I$/d$V$ map on the cleaved surface of FeGe were performed by adopting the same procedure as in Ref.[24].

### Single crystal x-ray diffraction and structural refinement

SXRD at Mo-K$_{\alpha}$ radiation were performed on a Bruker D8 Venture diffractometer. The mounted crystal (the batch annealed at 643 K) was heated up to 400 K under the $N_2$ gas flow. SXRD with synchrotron radiation was measured at the EH2 station of beamline P24 of PETRA-III at DESY in Germany. Pilatus CdTe 1M detector was used to collect the signal of the diffracted x-rays with a wavelength of 0.5 Å. The detailed measurement procedures and strategies can be found in Ref.[26] Complete datasets for structural refinements were collected successively at 270, 150, 110, 80, 60, 40, 20, 75 and 90 K. Continuous scans of 940 frames with an interval of 0.1 degree were performed during the cooling process at a rate of 1 K/min. The rotation speed of the goniometer head is 1 degree/second. The temperature for each data set was chosen to be the end point of the scan. Data reductions including integration and absorption correction for a complete dataset were performed by the software package Eval15[56] and SADABS[57]. The reconstructed reflection images in the reciprocal space and the integration of reflections for the temperature-dependent fast scan were completed by the software CrysAlis$^{pro}$ (CrysAlis Pro Version 171.40.67a, Rigaku Oxford Diffraction).

The (3+3)-dimensional commensurate modulated structures were refined using the software Jana2020[58,59] and assigning the commensurate sections $t_0$, $u_0$ and $v_0$ for the three independent vectors $q_1$ = (0.5, 0, 0), $q_2$ = (–0.5, 0.5, 0) and $q_3$ = (0, 0, 0.5), respectively. According to the table of all the possible superspace groups[60], those with a Laue group of $P6/mmm$ as the high temperature phase were considered first. The small values of $R_{int}$ for both satellite and main reflections corroborates the appropriate point symmetry for the reflections. $P6/mmm(\alpha_1,0,0)0000(-\alpha_1,\alpha_1,0)0000(0,0,\gamma_2)0000$ is the only plausible choice based on the systematic extinction condition of the reflections and the q-vectors. The initial phases ($t_0$, $u_0$, $v_0$) for three independent q-vectors determine the real structure for a commensurate modulated structure[53], which is an important step for the refinement. In order to preserve the six-fold rotation symmetry in the 2 × 2 × 2 supercell, the values of $t_0$ and $u_0$ should be zero. We varied the values of $v_0$ in the range (0, 0.5) and found out that $v_0$ = 0.25 gave the best $R$ values for the refinement, see Table S1. The regular space group in the 2 × 2 × 2 supercell for the commensurate sections ($t_0$, $u_0$, $v_0$) = (0, 0, 0.25) is $P6/mmm$, consistent with the recent report at 80 K[26]. The relative phases ($t$, $u$) for lattice translation operations in the *ab* plane are illustrated in Fig. S14. Both displacive and occupational modulation harmonic functions were added for specific combinations of q-vectors, see Supporting Information. The occupancy restrictions of Ge1_1 (in the Fe kagome plane) and Ge1_2 (out of the Fe kagome plane) atoms were also applied to match the mutually exclusive occupancy of these two sites in a real structure. The refinement results for 110, 80, 40, 20 and 400 K are summarized in Table S1-S16 of Supporting Information.

### Resonant Elastic x-ray Scattering (REXS)

Resonant elastic x-ray scattering experiments were performed at the BL02U2 surface diffraction beamline of the Shanghai Synchrotron Radiation Facility (SSRF) with a high-precision six-circle diffractometer. The sample surface size was 1 × 1 mm$^2$, and the x-ray beam spot size is 160 × 80 μm$^2$ (FWHM). The incident photon energy was set to the Ge K-edge resonance energy of 11.133 keV, determined from the fluorescence measurements on the samples. The diffraction signals were collected by an Eiger 500K pixel detector. The sample temperature was controlled using a helium closed-cycle cryostat. The FWHM of the peak is estimated by a Gaussian function fit after ignoring the shoulder signal from slightly misaligned crystals.

### Density Functional Theory (DFT) Calculations

Electronic structure calculations were performed within the density functional theory framework with projector-augmented wave potentials[61,62], using the Vienna ab initio simulation package (VASP)[63,64]. The Perdew-Burke-Ernzerhof parameterization of the generalized gradient approximation was employed to consider exchange-correlation effects[65]. An energy cutoff of 400 eV was set for the plane wave basis set, and a 6 × 6 × 8 Γ-centered k-point mesh was used for Brillouin zone sampling for the 2 × 2 × 2 superstructures. The spin-orbit coupling was included self-consistently to incorporate relativistic effects in the calculations. In the case of collinear antiferromagnetic configuration, an out-of-plane A-type spin arrangement was considered for the Fe atoms[66]. The canted configuration was modelled by introducing a rotation of the spins by a small degree with respect to the z-axis.

### Acknowledgements

The authors thanks Chenchao Xu, Zhaopeng Guo and Jun Li for their insightful discussions. J.-K. B. acknowledges support from the National Natural Science Foundation of China (Grant No. 12204298), Beijing National Laboratory for Condensed Matter Physics (Grant No. 2023BNLCMPKF019) and the startup funding of Hangzhou Normal University. S. X. C. would like to acknowledge the research grant from the National Natural Science Foundation of China (Grant No. 12374116). L. X. R. acknowledges support from the Ministry of Science and Technology of China (Grant No. 2022YFA1603900). J. X. Y. acknowledges the support from the National Key R&D Program of China (No. 2023YFA1407300) and the National Science Foundation of China (No. 12374060). Y. Q. would like to acknowledge the National Natural Science Foundation of China (Grant No. 52272265). Y. K. L. thanks the support of the Hangzhou Joint Fund of the Zhejiang Provincial Natural Science Foundation of China (under Grants No. LHZSZ24A040001). The research at the University of Bayreuth has been funded by the Deutsche Forschungsgemeinschaft (DFG, German Research Foundation) – 406658237. B. S. acknowledges the support from the Department of Atomic Energy, Government of India, under Project No. 12-R&D-TFR-5.10-0100, and the computational resources of TIFR Mumbai. We thank M. Tolkiehn and C. Paulmann for their assistance at Beamline P24, and Y. T. Song and L. M. Shao for their help in the lab-source x-ray diffraction. We acknowledge DESY (Hamburg, Germany), a member of the Helmholtz Association HGF, for the provision of experimental facilities. Parts of this research were carried out at PETRA III, using beamline P24. Beamtime was allocated for proposal I-20220188. We thank BL02U2 of Shanghai Synchrotron Radiation Facility for experiment beamtime.

## Additional information

**Supplementary information.** The online version contains supplementary material available.

## Data Availability

All the data are available upon request from the corresponding authors.

## Competing interests

The authors declare no competing interests.

## Author Contributions

J.-K.B. designed research, C.F.S., W.C.H., H.B.D, B.P., S.R.K., Y.L., S.R., C.E., H.A., L.N., J.-Y.L., T.Y.Y., G.W.L., B.B.M., Q.W., Z.D.L., B.J.K., W.T.Y., Y.C.L., Z.H.Y, Y. X. C., X. L., Y.K. L., Y.P.Q., A.T., W.R., G.-H.C., J.-X.Y., B.S., X.R.L., S.v.S., S.X.C., and J.K.B. performed research. C.F.S., W.C.H., J.-X.Y., X.R.L., B.B.M., S.R.K., S.R., B.S., B.P., S.v.S. and J.K.B. analyzed data. C.F.S., W.C.H., B. S., X.R.L., S.v.S. and J.K.B. wrote the paper.

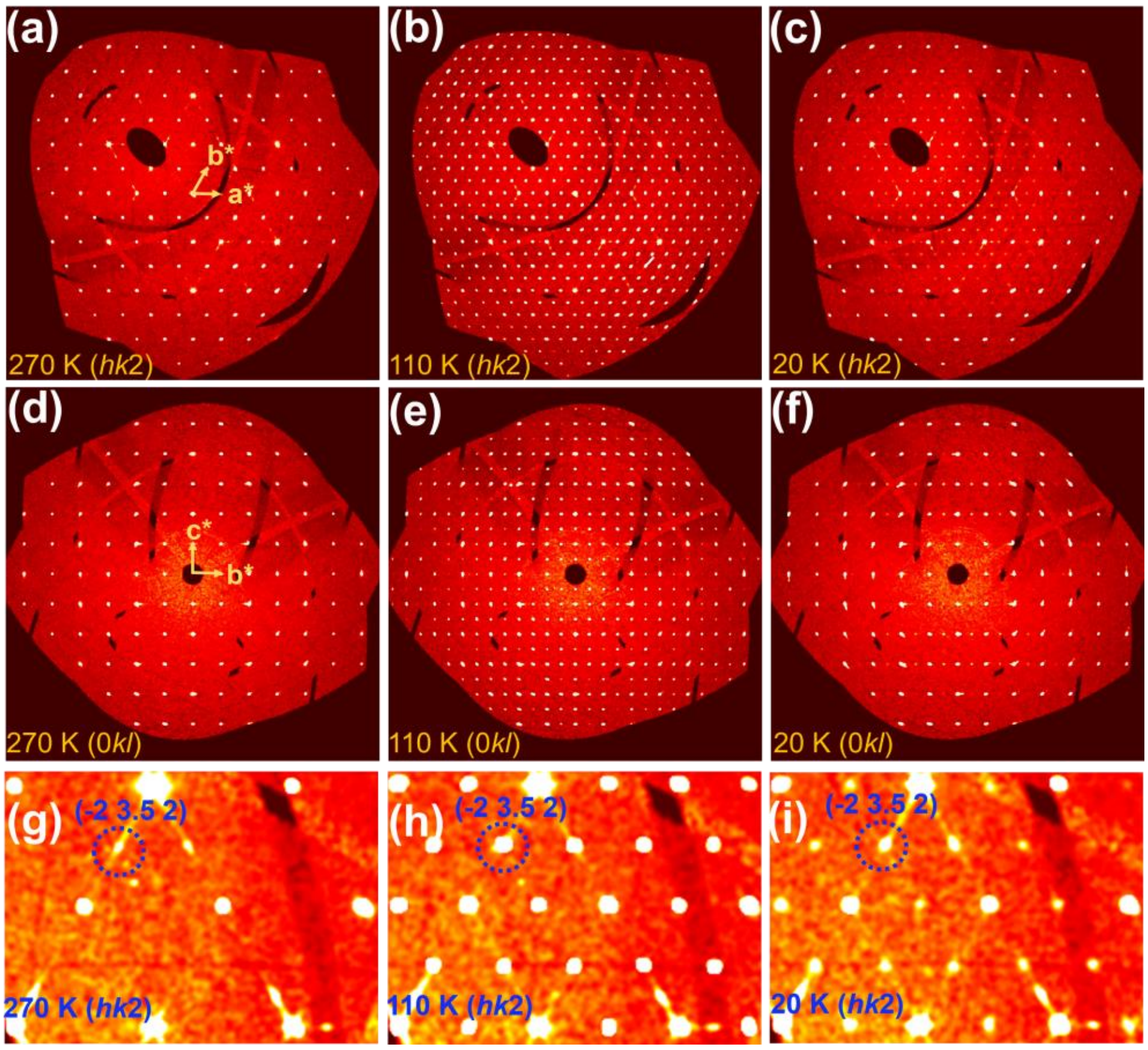

Fig. 1. Reconstructed images of reflections in the reciprocal space for the annealed sample. (a)-(c) (*hk*2) planes at 270, 110 and 20 K, respectively. (d)-(f) (0*kl*) planes at 270, 110 and 20 K, respectively. (g)-(i) The zoomed areas of the diffuse scattering for (*hk*2) planes at 270, 110 and 20 K, respectively. Some reflection indices were given. The data at 270 K are reproduced from the reference[26].

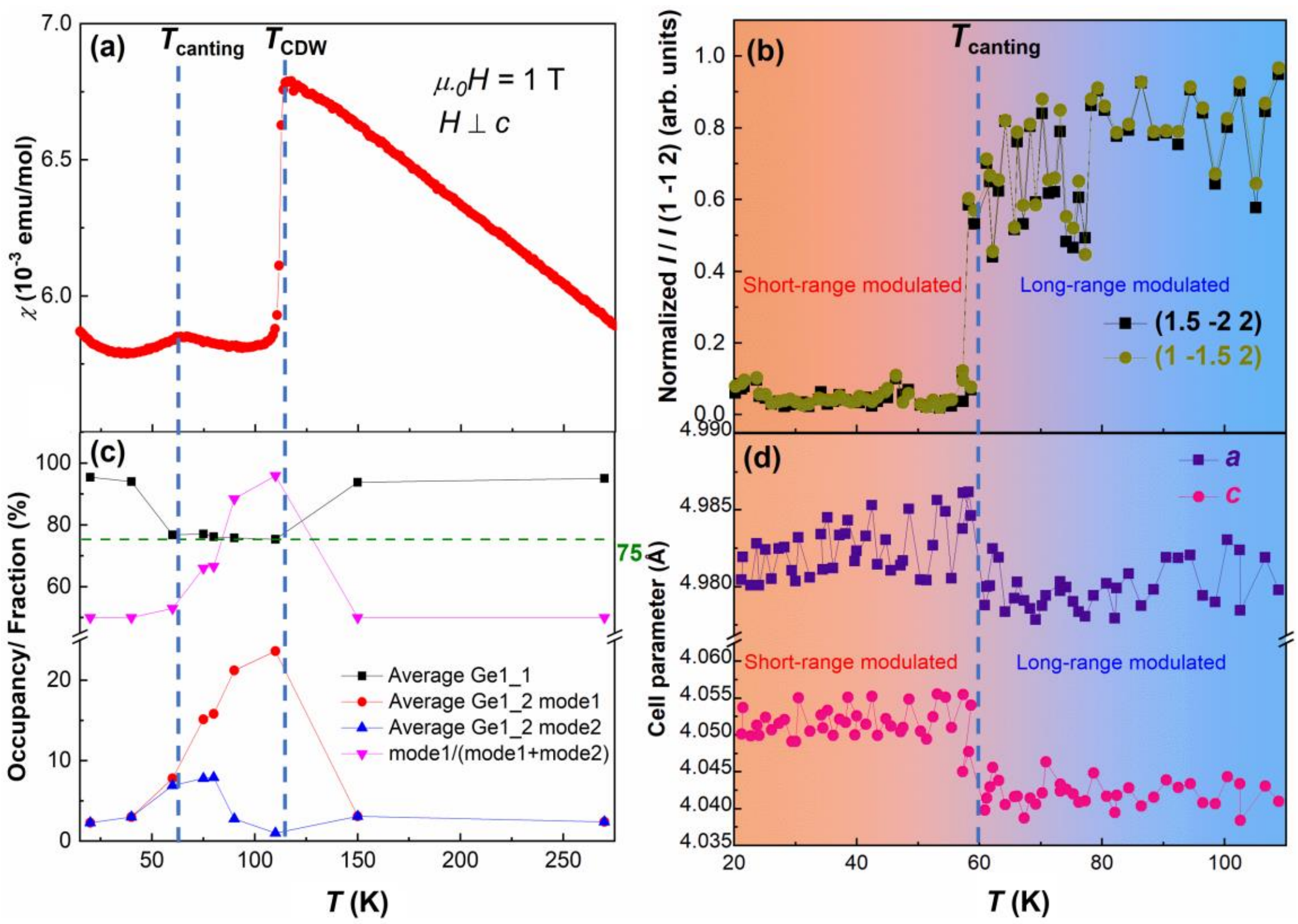

Fig. 2. Temperature-dependent magnetic susceptibility, superlattice peak intensity and lattice parameters. (a) Magnetic susceptibility under $H \perp c$ ($\mu_0 H$ = 1 T) in FeGe. (b) Normalized intensities for superlattice peaks (1.5 –2 2) and (1 –1.5 2). (c) Average occupancies of Ge1_1 site in the kagome plane and Ge1_2 site originating from two different distortion modes along the *c* axis, and the fraction of the dominant mode 1. (d) Lattice parameters of *a* and *c*.

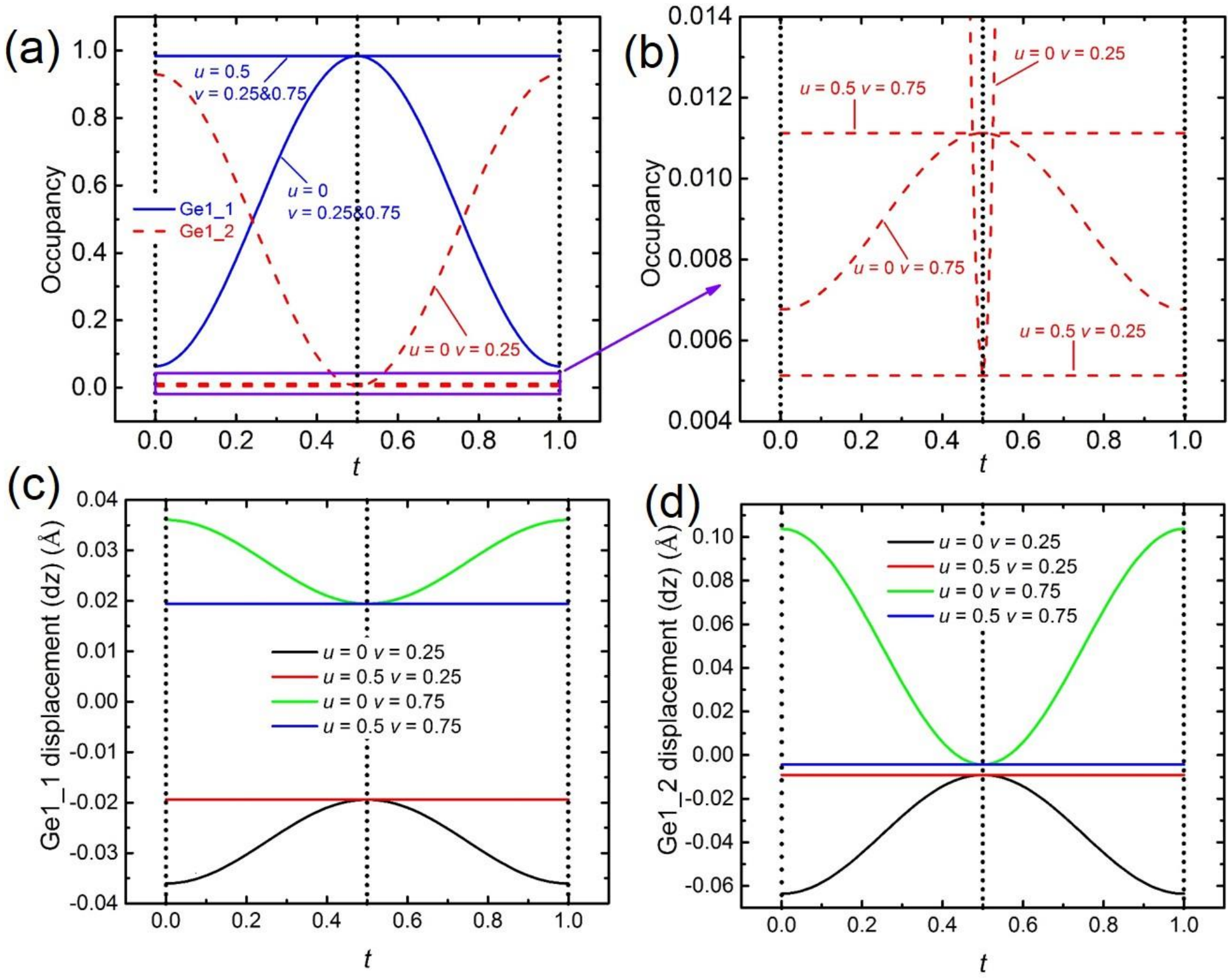

Fig. 3. The *t*-plots of occupational and displacive modulations at Ge1 sites at 110 K for the annealed crystals in the commensurate section cuts (*u*, *v*). (a) The occupational modulations of Ge1_1 (blue solid) and Ge1_2 (red dash) atoms in four commensurate section cuts (*u*, *v*) at 110 K for the annealed crystal: (0, 0.25), (0.5, 0.25), (0, 0.75) and (0.25, 0.75). (b) The zoomed rectangular area for the modulation information of Ge1_2 atoms in (a). (c) and (d) The displacive modulations of Ge1_1 and Ge1_2 atoms along *z* direction in the same commensurate cuts, respectively. The dashed vertical lines are the commensurate section cuts for a real structure.

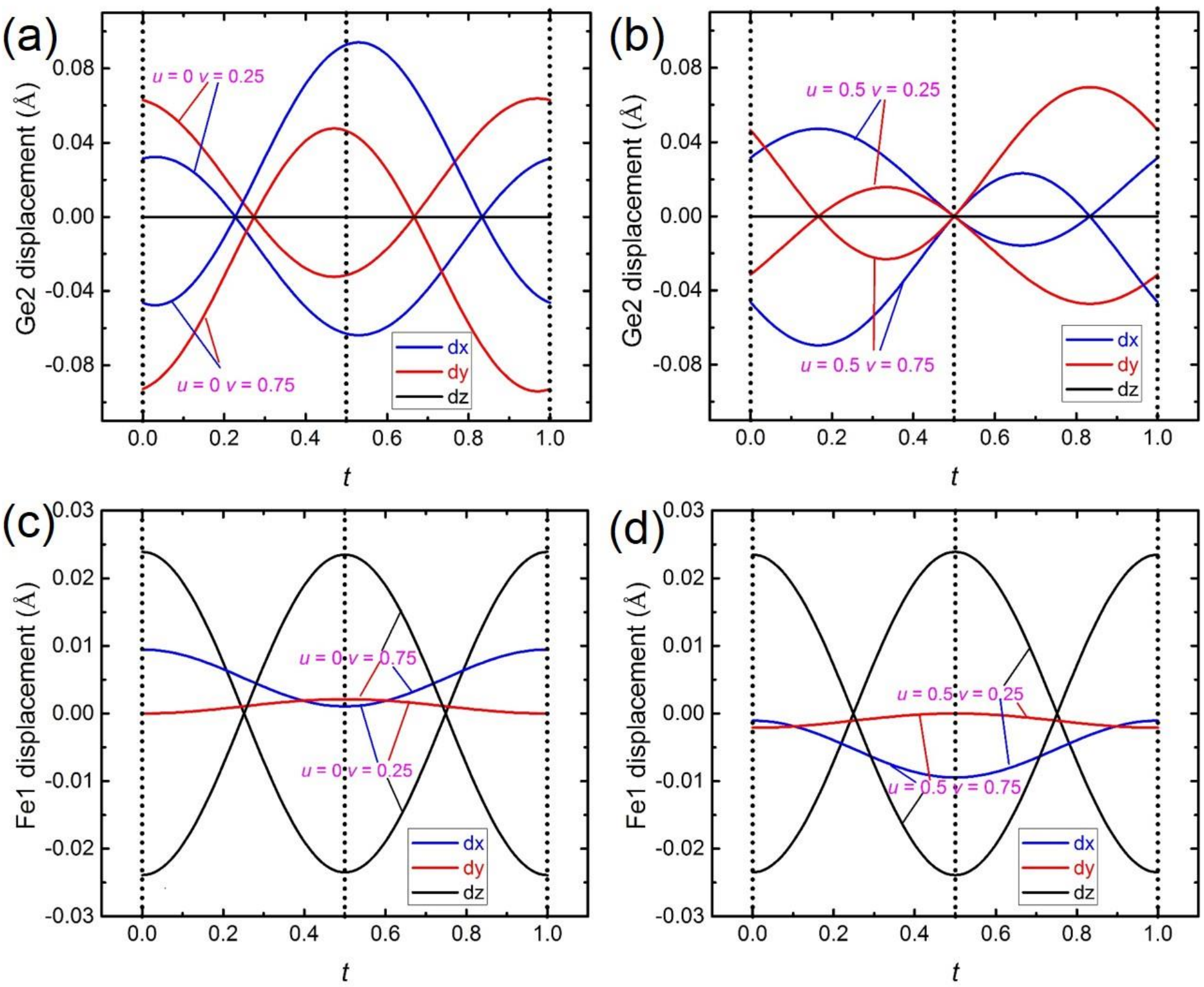

Fig. 4. *t*-plots of displacive modulations along *x*, *y* and *z* directions at 110 K for the annealed crystals in the commensurate section cuts (*u*, *v*). (a) (0, 0.25) and (0, 0.75) for Ge2 atom, (b) (0.5, 0.25) and (0.5, 0.75) for Ge2 atom, (c) (0, 0.25) and (0, 0.75) for Fe1 atom and (d) (0.5, 0.25) and (0.5, 0.75) for Fe1 atom. The dashed vertical lines are the commensurate section cuts for the actual structure.

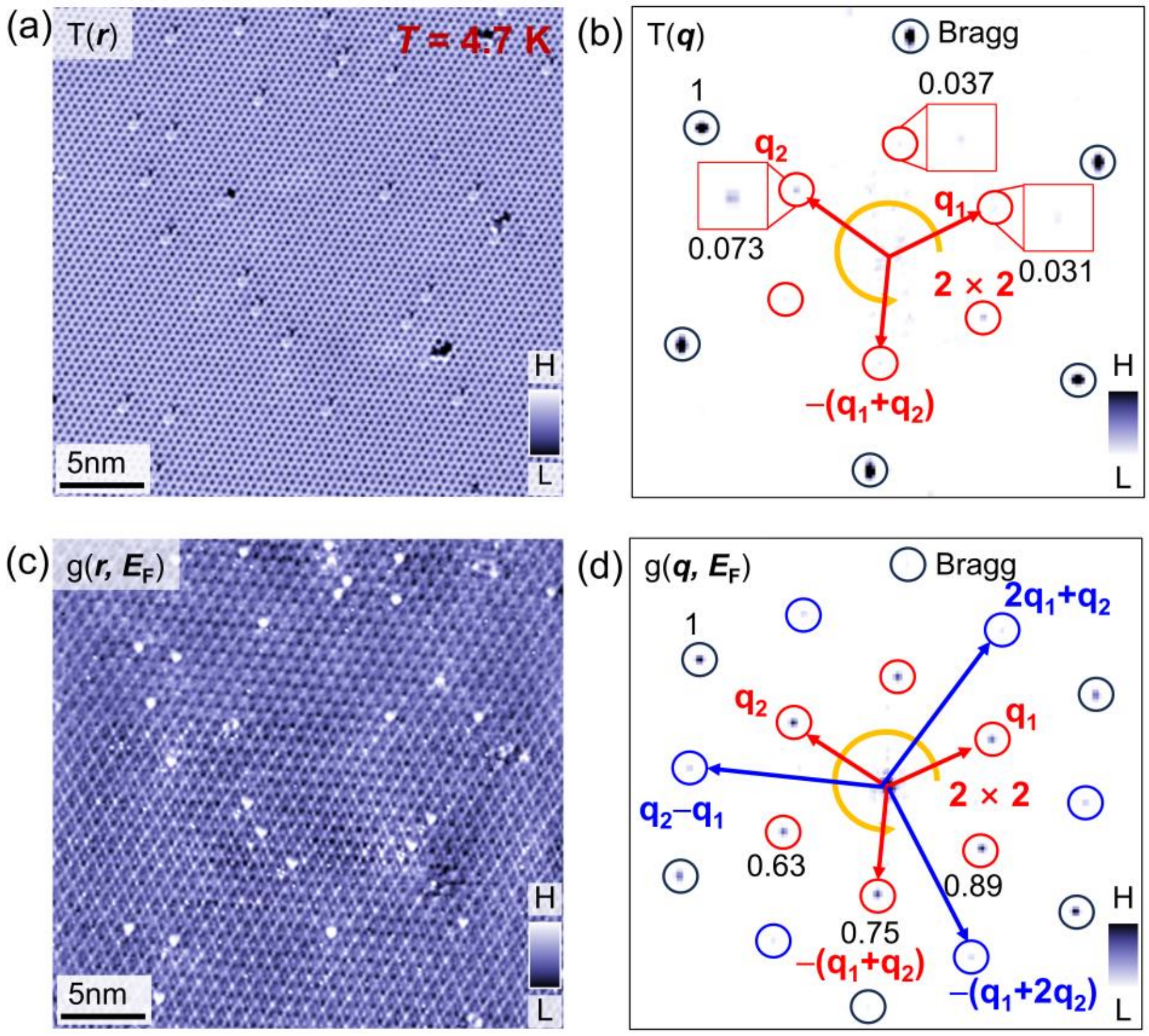

Fig. 5. The STM measurement results for annealed crystal of FeGe. (a) Topographic image of STM and (b) its corresponding Fourier transform at 4.7 K in FeGe. The normalized intensities of different q-vector superlattice reflections are marked by the numbers (c) Differential conductance d*I*/d*V* map taken at Fermi level at 4.7 K and (d) its corresponding Fourier transform. The Bragg and CDW peaks were marked with black and red (blue) circles, respectively. The **q-**vectors for all the CDW peaks are also labeled by blue and red arrows. The normalized intensities of different q-vector CDW peaks are marked by the numbers. The chirality of the CDW is marked by the yellow arrows.

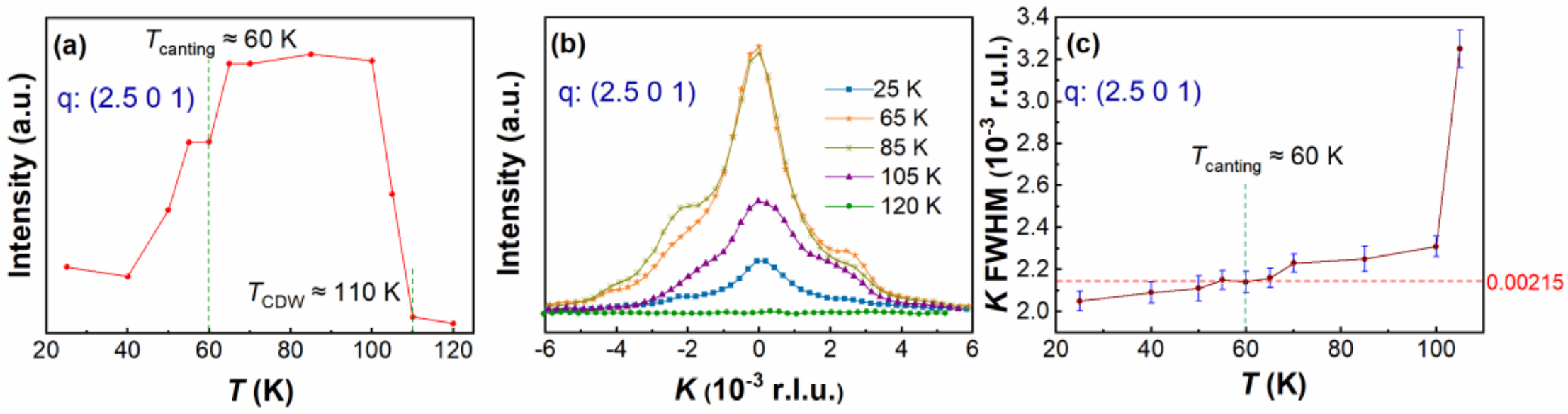

Fig. 6. The REXS measurement results at the moment transfer **q** = (2.5 0 1). (a) Temperature-dependent intensity of the peak with a moment transfer **q** = (2.5 0 1), (b) *K* scan of the peak at different temperatures. (c) Temperature-dependent FWHM of the peak **q** in (b) with the error marked.

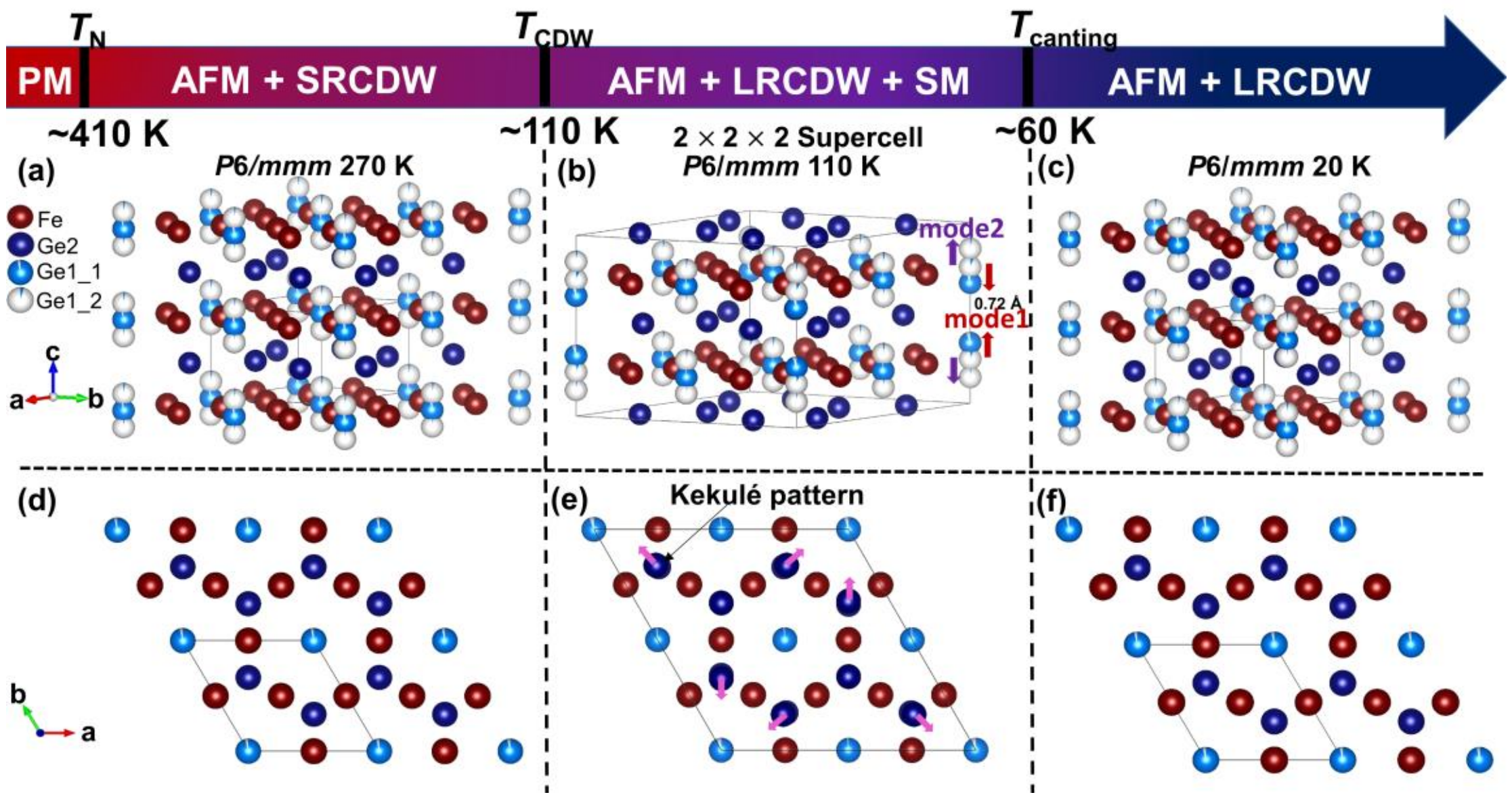

Fig. 7. Temperature-dependent quantum states and crystal structures for annealed FeGe crystals. (a)-(c) The crystal structure models of annealed FeGe crystal at 270, 110 K and 20 K correspond to the short-range, long-range and short-range structural modulation models, respectively. The dimerization of Ge1 atoms in the kagome plane along the *c* axis for mode1 and mode2 are marked by red and purple arrows, respectively. (d)-(f) The crystal structure models viewed along the *c* axis at 270, 110 K and 20 K, respectively. AFM, SRCDW, LRCDW and SM are abbreviated from antiferromagnetism, short-range charge density wave, long-range charge density wave and structural modulation, respectively.